\documentclass[twocolumn,showpacs,preprintnumbers,superscriptaddress,amsmath,amssymb]{revtex4-1}

\usepackage{epsfig}
\usepackage{dcolumn}
\usepackage{bm}
\usepackage{times}
\usepackage{xcolor}
\usepackage{amssymb}
\usepackage{amsthm}
\usepackage{graphicx}
\usepackage{multirow}
\usepackage{subfigure}
\usepackage{booktabs}
\usepackage{url}
\usepackage{lineno}
\usepackage{algorithm, algorithmic}

\begin{document}


\title{On Predictability of Time Series}

\author{Paiheng Xu}
\affiliation{CompleX Lab, University of Electronic Science and Technology of China, Chengdu 611731, People's Republic of China}
\affiliation{School of Hanhong, Southwest University, Chongqing 400715, People's Republic of China}

\author{Likang Yin}
\affiliation{CompleX Lab, University of Electronic Science and Technology of China, Chengdu 611731, People's Republic of China}
\affiliation{School of Hanhong, Southwest University, Chongqing 400715, People's Republic of China}

\author{Zhongtao Yue}
\affiliation{CompleX Lab, University of Electronic Science and Technology of China, Chengdu 611731, People's Republic of China}
\affiliation{Big Data Research Center, University of Electronic Science and Technology of China, Chengdu 611731, People's Republic of China}
\affiliation{Institute of Fundamental and Frontier Sciences,University of Electronic Science and Technology of China, Chengdu 611731, People's Republic of China}

\author{Tao Zhou}\email{zhutou@ustc.edu}
\affiliation{CompleX Lab, University of Electronic Science and Technology of China, Chengdu 611731, People's Republic of China}
\affiliation{Big Data Research Center, University of Electronic Science and Technology of China, Chengdu 611731, People's Republic of China}

\date{\today}

\begin{abstract}

The method to estimate the predictability of human mobility was proposed in [C. Song \emph{et al.}, Science {\bf 327}, 1018 (2010)], which is extensively followed in exploring the predictability of disparate time series. However, the ambiguous description in the original paper leads to some misunderstandings, including the inconsistent logarithm bases in the entropy estimator and the entropy-predictability-conversion equation, as well as the details in the calculation of the Lempel-Ziv estimator, which further results in remarkably overestimated predictability. This paper demonstrates the degree of overestimation by four different types of theoretically generated time series and an empirical data set, and shows the intrinsic deviation of the Lempel-Ziv estimator for highly random time series. This work provides a clear picture on this issue and thus helps researchers in correctly estimating the predictability of time series.

\end{abstract}



\maketitle

\section{Introduction} \label{sec.intro}

In recent years, the rapidly increasing usages of Global Positioning System (GPS), ranging from mobile phones, fitness bracelets and vehicle positioning systems, provide us with unprecedentedly rich information to capture and analyze human mobility patterns. As a result, the prediction of human mobility has received growing attention for its importance in traffic management \cite{wang2015predictability,ren2013potential,li2014limits,zhao2016predicting}, disaster response \cite{lu2012predictability,bengtsson2011improved,kenett2012population}, epidemic prevention \cite{stoddard2009role,belik2011natural,barmak2011dengue}, and so on \cite{barbosa2018human}.

\par Though the prediction is getting more and more accurate due to advanced algorithms and the availability of vast data, it is not clear how well these algorithms perform versus the best possible prediction. Accordingly, the predictability of human mobility was proposed, which aims at measuring the theoretically maximum prediction accuracy $\Pi_{max}$ for the given data. Song \emph{et al.} proposed an entropic framework \cite{song2010limits} to calculate the predictability $\Pi_{max}$ by solving a limited case of the Fano inequality \cite{fano1961transmission,brabazon2008natural}. Empirical analysis on this basis \cite{song2010limits} suggested that the predictability of human mobility could reach $93\%$ on average.

\par Many scientists tried to design advanced predicting algorithms that can approach the predictability \cite{lian2015mining}, such as the Markovian model \cite{lu2013approaching}, the neural network algorithm \cite{mukai2012taxi}, the sequence-based model \cite{zhao2016predicting}, and so on. The information from social networks \cite{ponieman2013human,jurdak2015understanding}, semantic labels \cite{ying2011semantic}, demographic characteristics \cite{yang2017indigenization}, and activity patterns \cite{yu2017modeling} are also utilized to improve the predicting algorithms.

\par Another interesting topic is to dig out significant factors that affect the predictability, such as the temporal and spatial resolution of data \cite{lin2012predictability,cuttone2018understanding,jensen2010estimating,yan2014universal,yan2017universal}, the preference of exploration \cite{pappalardo2015returners,cuttone2018understanding} and the data quality \cite{iovan2013moving}. In particular, the predictability of next-timestep prediction is shown to have a very high upper bound ($>90\%$), which is mostly due to the stationarity in the human mobility \cite{Yan2013,ikanovic2017alternative}. This makes the next-place prediction a more challenging and attractive topic.

\par Some other researchers have tried different methods or modified entropic measures to quantify the predictability of human mobility, such as mutual information \cite{takaguchi2011predictability,chen2016temporal}, instantaneous entropy \cite{baumann2013use,mcinerney2012exploring}, a contextual model which allows predictability to be assessed as the accuracy of the model in making predictions \cite{austin2014regularity}, and so on. Smith \emph{et al.} \cite{smith2014refined} integrated real-world topological constraints into the calculation of the upper bound and presented a refined predictability of human mobility. To remove the stationarity in human mobility, Ikanovic and Mollgaard \cite{ikanovic2017alternative} aimed at the next-place prediction and proposed an alternative approach independent of the temporal scale. Yao \emph{et al.} \cite{yao2004measure} proposed forecast entropy to measure the difficulty of predicting an observed time series based on the distributions of the time series in different spaces, and showed that the forecast entropy of a random system is clearly different from that of a deterministically chaotic system.

\par The analytical framework of predictability is also extended to other types of time series, such as human communication sequences \cite{takaguchi2011predictability,zhang2014analysis,Zhao2013}, vehicular mobility \cite{wang2015predictability,ren2013potential, li2014limits,zhao2016predicting,xu2017entropy}, the IP address sequence of cyberattacks \cite{chen2015spatiotemporal}, stock price change \cite{fiedor2014frequency}, consumer visitation pattern \cite{krumme2013predictability,Zhao2013}, online user behaviors \cite{jurdak2015understanding,chen2016temporal,sinatra2014entropy}, electronic health records \cite{dahlem2015predictability}, and so on.

\par However, the ambiguous description in Ref. \cite{song2010limits} leads to some misunderstandings, including the inconsistent logarithm bases in the entropy estimator and the entropy-predictability-conversion (EPC) equation, as well as the details in the calculation of the Lempel-Ziv (LZ) estimator. These misunderstandings will result in remarkably higher predictability than the true value. In this paper, we generate four types of artificial time series with controllable predictability, namely exploration sequence, random sequence, deterministic sequence and Markovian sequence, based on which we can quantify the deviation of theoretical estimation from the true value. We compare different possible understandings of the theoretical estimators in the literature, and clearly show the advantage of the right implementation, which is also consistent with previous theoretical entropy analyses on time series \cite{kontoyiannis1998nonparametric,grassberger1989estimating,ziv1978compression}.

Many researchers have followed Ref. \cite{song2010limits} to calculate the predictability of time series. Ciobanu \emph{et al.} \cite{ciobanu2014interaction} studied mobile interactions collected at the Politehnica University of Bucharest and explored its predictability in the opportunistic networks. Li \emph{et al.} \cite{li2014limits} proposed an areas transition model to describe the vehicular mobility among the areas divided by the city intersections and examined the predictabilities of large-scale urban vehicular networks. Zhao \emph{et al.} \cite{zhao2016predicting} obtained the theoretical predictability by entropy measure and used it to identify the effectiveness of different predicting algorithms. Xu \emph{et al.} \cite{xu2017entropy} defined travel time predictability as the probability to correctly predict the travel time by employing multiscale entropy. The predictability $\Pi_{max}$ obtained in the above-mentioned works may be higher than it supposed to be due to the unmatched choice of logarithm bases \cite{barbosa2018human}. The present paper could refine our knowledge in the above-mentioned issues. This paper also raises a challenge about how to accurately estimate predictability for less-predictable time series since the LZ estimator fails in such case. The four types of artificially generated time series can also be treated as the touchstone for the validity of newly proposed methods in the future.

\section{Theoretical Analysis} \label{sec.theoretical}

Considering a historical sequence ${T} = \left\{{{X_1},{X_2},\cdots,{X_n}} \right\}$ (other time series with finite and discrete values of elements can be treated in the same way), Song \emph{et al.} \cite{song2010limits} adopted the actual entropy,
\begin{equation}
S =  - \sum\nolimits_{{T}^\prime  \subset {T}} {P({T}^\prime ){{\log }_2}[P({T}^\prime )]},
\end{equation}
to measure the information capacity of such sequence, where $P\left( {{T}^\prime } \right)$ represents the probability of finding a subsequence ${{T}^\prime} $ in the trajectory $T$. Based on the Fano inequality \cite{fano1961transmission,brabazon2008natural}, they obtained the upper bound of the predictability ${\Pi _{\max }}$ by solving the following EPC equation
\begin{equation}\label{Eq.Fano}
\begin{split}
   H =
   &- \left[ {{\Pi ^{\max }}{{\log }_2}{\Pi ^{\max }} + (1 - {\Pi ^{\max }}){{\log }_2}(1 - {\Pi ^{\max }})} \right]\\
   & + (1 - {\Pi ^{\max }}){\log _2}(m - 1),
\end{split}
\end{equation}
where $m$ denotes the number of distinct locations appeared in $T$, and $H$ is the entropy rate of $T$, mathematically defined as
\begin{equation}
H=\lim_{n\rightarrow \infty} S(X_1,X_2,\cdots,X_n).
\end{equation}

\par The direct computation of actual entropy is highly time-consuming and thus usually infeasible for real time series, therefore an estimator for actual entropy based on the LZ data compression method \cite{kontoyiannis1998nonparametric,grassberger1989estimating} is applied. For a time series with length $n$, the entropy is estimated by
\begin{equation}\label{Eq.est}
  {S^{^{est}}} = {\left( {{\textstyle{1 \over n}}\sum\limits_i {{\Lambda _i}} } \right)^{ - 1}}\ln n,
\end{equation}
where ${\Lambda_i}$ denotes the minimum length $k$ such that the sub-sequence starting from position $i$ with length $k$ does not appear as a continuous sub-sequence of $\{X_1,X_2,\cdots,X_{i-1}\}$.

The above description, extracted from Ref. \cite{song2010limits}, is ambiguous in two aspects. Firstly, the logarithm base in Eq. (4) is not clarified and thus was usually taken as the Euler's constant, namely $e\approx 2.7183$ \cite{barbosa2018human}. Secondly, when every sub-sequence starting from $X_i$ appears as a sub-sequence of $\{X_1,X_2,\cdots,X_{i-1}\}$, how to determine $\Lambda_i$ is a puzzle.

\begin{figure}
  \centering
  \includegraphics[width=1\linewidth]{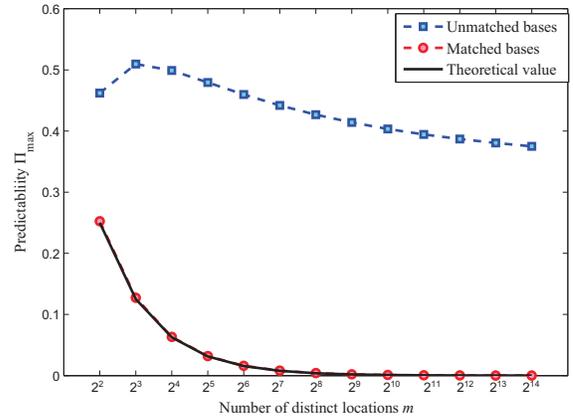}\\
  \caption{(Color online) The predictability with unmatched bases (Eq. (4), blue squares) and matched bases (Eq. (5), red circles), compared with the theoretical value (black line) for exploration sequences with varying $m$. The value of $\Lambda_i$ is set to be $n-i+2$ if every sub-sequence starting from $X_i$ appears as a sub-sequence of $\{X_1,X_2,\cdots,X_{i-1}\}$.}\label{fig.exploration.pre}
\end{figure}

To quantify the accuracies of estimated predictability of different implementations, we consider four theoretical generators of time series with controllable predictability. Supposing there are $m$ distinct locations in the constructed time series $T$ with length $n$, the four types of sequences are as follows. (i) Exploration sequence. We set $m=n$ and generate a random permutation with the $m$ elements, so that every step in $T$ can be considered as an exploration based on the previous historical trajectory. Since we do NOT know the information that the next location is a new location, the theoretical predictability should be $1/m$. (ii) Random sequence. Every elements in $T$ are independently and randomly generated and thus the theoretical value should be $1/m$ when $n$ approaches infinity. (iii) Deterministic sequence. Without loss of generality, the constructed deterministic time series is $T=\{X_1,X_2,\cdots,X_m,X_1,X_2,\cdots\}$, whose predictability should converge to 1 when $n$ approaches infinity. (iv) Markovian sequence. At each step, with probability $p$, the next location is determined by the same path as the deterministic sequence, say ${X_1}\rightarrow{X_2}\rightarrow\cdots\rightarrow{X_m}\rightarrow{X_1}\rightarrow\cdots$, and with probability $1-p$, the next location is randomly selected from the $m$ candidates. Accordingly, the theoretical predictability should be $p+(1-p)/m$ when $n$ approaches infinity.

We first consider the deviation caused by the unmatched logarithm bases in Eq. (1) and Eq. (4), as described in Ref. \cite{song2010limits}. According to the previous literature \cite{kontoyiannis1998nonparametric,grassberger1989estimating,ziv1978compression}, the two bases should be the same. In particular, Grassberger \cite{grassberger1989estimating} suggested to take the logarithm to base $2$ in order to obtain $S^{est}$ in bits. Therefore, we replace Eq. (4) by
\begin{equation}\label{Eq.est.revised}
{S^{^{est}}} = {\left( {{\textstyle{1 \over n}}\sum\limits_i {{\Lambda _i}} } \right)^{ - 1}}\log_2 n
\end{equation}
to obtain the matched case (to replace $\log_2$ in Eq. (4) by $\ln$ will generate the same result). We first compare the two different cases based on the exploration sequences. As shown in Fig. 1, for the unmatched case, the predictability ${\Pi _{\max }}$ is larger than 0.35 even when $m$ is $2^{14}$, while the theoretical value $1/m$ should be already very close to zero. At the same time, ${\Pi _{\max }}$ obtained in the matched case (Eq. (5)) is in accordance with the theoretical value.

\begin{figure}
  \centering
  \subfigure{
    \label{fig.random.10} 
    \includegraphics[width=1\linewidth]{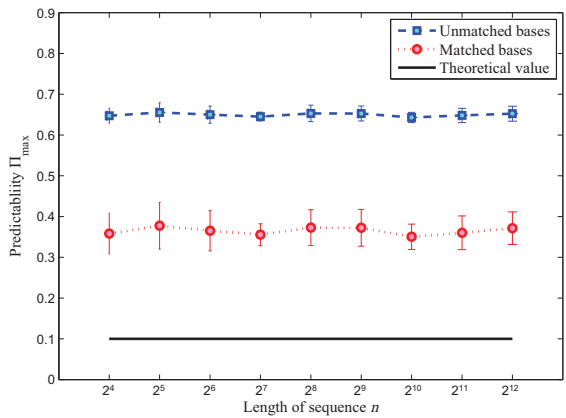}}
  \caption{(Color online) The predictability with unmatched bases (Eq. (4), blue squares) and matched bases (Eq. (5), red circles), compared with the theoretical value (black line) for random sequences with varying $n$. The number of distinct locations $m$ is fixed to be 10 (for other values of $m$ $(<n)$, the results are the same). The simulation results are obtained by 100 independent implementations, with error bars denote the standard deviations. The value of $\Lambda_i$ is set to be $n-i+2$ if every sub-sequence starting from $X_i$ appears as a sub-sequence of $\{X_1,X_2,\cdots,X_{i-1}\}$.}\label{fig.random}
\end{figure}

We further compare the matched and unmatched cases for random and deterministic sequences. Figure 2 shows typical results with varying $n$, where one can observe that both matched and unmatched estimators are deviated from the theoretical value, while the matched case performs relatively better. In the other extreme relative to the random sequences, for deterministic sequences, when $n$ becomes much larger than $m$, both estimators converges to the theoretical value 1 quickly, as shown in Figure 3. Figure 4 reports the results for Markovian sequences, whose predictability can be precisely controlled by adjusting the parameter $p$, with $p=0$ and $p=1$ corresponding to the two extremes, namely random sequences and deterministic sequences, respectively. Comparing the two curves with the same setting of $\Lambda_i$ as Fig. 1 to Fig. 3 (marked as $\Lambda=k+1$ in Fig. 4), one can observe three phenomena: (i) the estimated predictability of unmatched case is always higher than that of matched case, (ii) the predictability of matched case is overall closer to the theoretical value than that of unmatched case, (iii) the deviation of LZ estimator for highly random sequences (i.e., small $p$) is remarkably higher than that for more predictable sequences (i.e., large $p$).

\begin{figure}
  \centering
  \subfigure{
    \includegraphics[width=1\linewidth]{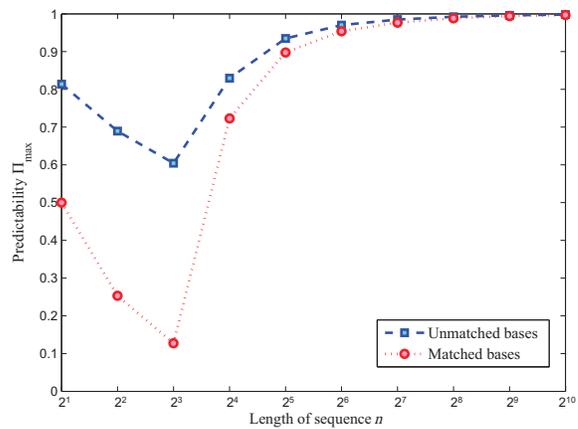}}
  \caption{(Color online) The predictability with unmatched bases (Eq. (4), blue squares) and matched bases (Eq. (5), red circles) for deterministic sequences with varying $n$. The number of distinct locations $m$ is fixed to be 10. The value of $\Lambda_i$ is set to be $n-i+2$ if every sub-sequence starting from $X_i$ appears as a sub-sequence of $\{X_1,X_2,\cdots,X_{i-1}\}$.}\label{fig.deterministic}
\end{figure}

Another aspect we would like to clarify is how to determine the value of $\Lambda_i$ when every sub-sequence starting from $X_i$ appears as a sub-sequence of $\{X_1,X_2,\cdots,X_{i-1}\}$. A straightforward treatment is to set $\Lambda_i$ as
\begin{equation}
\Lambda_i=n+1.
\end{equation}
Let's look closely into the definition of $\Lambda_i$. ${\Lambda_i}$ is the minimum length $k$ such that the sub-sequence starting from position $i$ with length $k$ does not appear as a continuous sub-sequence of $\{X_1,X_2,\cdots,X_{i-1}\}$, which can also be explained as one plus the length $k^{(i)}_{\max}$ of the longest sub-sequence starting from position $i$ that appears as a continuous sub-sequence of $\{X_1,X_2,\cdots,X_{i-1}\}$, say
\begin{equation}
\Lambda_i=k^{(i)}_{\max}+1.
\end{equation}
Notice that, Eq. (7) is a unified explanation that can also be applied in the case when every sub-sequence starting from $X_i$ appears as a sub-sequence of $\{X_1,X_2,\cdots,X_{i-1}\}$, where $k^{(i)}_{\max}=n-i+1$ and thus $\Lambda_i=n-i+2$. As shown in Fig. 4, estimator based on Eq. (7) performs much better than that based on Eq. (6). Some other possible alternatives of the understanding of $\Lambda_i$ when every sub-sequence starting from $X_i$ appears as a sub-sequence of $\{X_1,X_2,\cdots,X_{i-1}\}$, such as $\Lambda_i=0$ and $\Lambda_i=n$ performs no better or even worse than Eq. (6). So we can conclude that Eq. (7) is an proper and unified understanding of $\Lambda_i$, and indeed it has been applied in Figs. 1 to 3.

\begin{figure}
  \centering
  \subfigure{
    \label{fig.markov.5} 
    \includegraphics[width=1\linewidth]{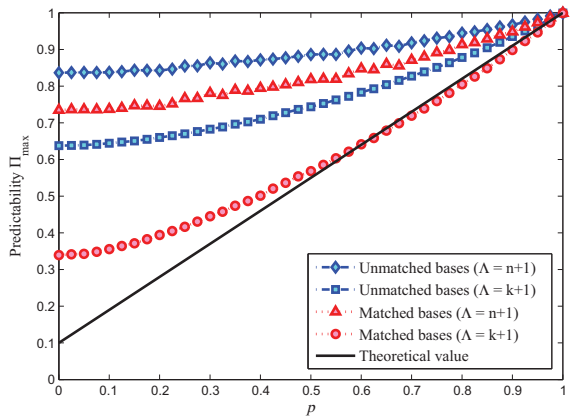}}
  \caption{(Color online) The predictability with unmatched bases (Eq. (4), in blue) and matched bases (Eq. (5), in red) for Markovian sequences with different $p$. The number of distinct locations is fixed as $m=10$ and the length of sequence is fixed as $n=2^{10}$. We also test different understandings of $\Lambda_i$ when every sub-sequence starting from $X_i$ appears as a sub-sequence of $\{X_1,X_2,\cdots,X_{i-1}\}$, which are respectively marked as $\Lambda=n+1$ (Eq. (6)) and $\Lambda=k+1$ (Eq. (7)) in the plot. The simulation results are obtained by averaging over 100 independent implementations. }\label{fig.markov.new}
\end{figure}

\section{Empirical Analysis} \label{sec.case}

This section shows the difference between predictabilities estimated with unmatched and matched logarithm bases by a real data set recording interaction traces among 66 participants in the Politehnica University of Bucharest during March to May 2012 \cite{marin2012exploring,ciobanu2014interaction}. In the experiments, each participant carries an Android smartphone with tracing function that can identify other participants if they are close enough (by Bluetooth or AllJoyn). Therefore, we obtain a sequence of interacting persons for each participant. The original sequences describe next-timestep interactions and since the updates are frequency, they are many continuous sub-sequences consisting of the same interacting person. Therefore, we compress such a sub-sequence into only one element. For example, if a participant A's original interaction sequence is BBBBCCCDCCCCCBB, it will be transformed into BCDCB.

After the above pretreatment, we remove all participants with sequence lengths no more than 5 and a few participants whose entropy rates do NOT converge according to Eq. (3) (also because of too small $n$ compared with $m$). Figure 5 reports the estimated predictabilities of the remain 22 participants. In accordance with the theoretical analysis, for each participant, the estimated predictability by unmatched logarithm bases is always remarkably larger than that by matched bases, and the averaged values of $\Pi_{\max}$ for the two cases are 0.63 and 0.39, respectively.

\begin{figure}
  \centering
  \includegraphics[width=1\linewidth]{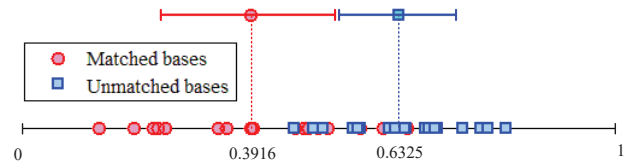}\\
  \caption{(Color online) Predictability distributions for the 22 valid samples, respectively obtained by applying unmatched bases (Eq. (4), blue squares) and matched bases (Eq. (5), red circles). $\Lambda_i$ is determined by Eq. (7). The blue and red horizontal lines denote the corresponding error bars (i.e., standard deviations).}\label{fig.predictability.case}
\end{figure}

\section{Conclusions} \label{sec:conclusion}

This paper briefly reviewed the framework proposed in Ref. \cite{song2010limits} for quantifying the predictability of human mobility. The ambiguous description in Ref. \cite{song2010limits} may lead to different understandings in some calculation details. We introduce some possible understandings, of which all the incorrect ones will result in overestimated predictability. When applying the considered method on human mobility or extending it to other discrete time series, we provide two clear suggestions. Firstly, the logarithm bases in the entropy estimator and the EPC equation should be the same. Secondly, $\Lambda_i$ should be explained in a clear and unified way as one plus the length of the longest sub-sequence starting from position $i$ that appears as a continuous sub-sequence of $\{X_1,X_2,\cdots,X_{i-1}\}$.

Theoretical analysis on time series with controlled predictability showed that the estimator proposed in Ref. \cite{song2010limits} failed when the time series is highly random. Therefore, at the end of this paper, we raise an open challenge for future study, that is, how to accurately estimate predictability for such less-predictable time series.

\section*{Acknowledgement}

The reported deviation from the true value was firstly found by Xiaoyong Yan and Zimo Yang. The authors would like to thank Zehui Qu for helpful discussion. This work was partially supported by the National Natural Science Foundation of China (Grants $61433014$, $61673085$, and $61603074$).


\end{document}